\documentclass[english
]{aa}
\usepackage{epsf,amsfonts,amssymb,graphicx,fancyheadings,caption}
\usepackage{natbib}
\usepackage{babel}
\usepackage{txfonts}
\usepackage{float} 
\bibpunct{(}{)}{;}{a}{}{,}

\begin{document}

\title{New R Coronae Borealis stars discovered in OGLE-III Galactic bulge fields from their mid- and near- infrared properties
}

\author{
P.~Tisserand\inst{1},
L.~Wyrzykowski\inst{2,3},
P.R.~Wood\inst{1},
A.~Udalski\inst{3},
M.K.~Szyma{\'n}ski\inst{3},
M.~Kubiak\inst{3},
G.~Pietrzy{\'n}ski\inst{3,4},
I.~Soszy{\'n}ski\inst{3},
O.~Szewczyk\inst{3,4},
K.~Ulaczyk\inst{3},
R.~Poleski\inst{3}
}

\institute{
Research School of Astronomy and Astrophysics, Australian National University, Cotter Rd, Weston Creek
 ACT 2611, Australia \and
Institute of Astronomy, University of Cambridge, Madingley Road, Cambridge CB3 0HA, England \and
Warsaw University Astronomical Observatory, Al. Ujazdowskie 4, 00-478 Warszawa, Poland \and
Universidad de Concepci{\'o}n, Departamento de Fisica, Astronomy Group, Casilla 160-C, Concepci{\'o}n, Chile
}

\offprints{P. Tisserand; \email{tisserand@mso.anu.edu.au}}

\date{Received ; Accepted}

%
%

\abstract {An R Coronae Borealis (RCB) star is a rare type of supergiant star that is increasingly suspected to be the evolved merger product of two white dwarfs. Lately, many RCBs have been found distributed in a thin disk structure embedded inside the Galactic bulge. This unexpectedly high density may provide additional insight into the nature and age of RCB stars.}
{We apply and test a new technique to find RCB stars based on their particular infrared emission due to circumstellar shell. We attempt to demonstrate that RCB stars can be identified without performing a light curve analysis, which would simplified the search outside optically monitored fields.} 
{We selected RCB candidates based on their near-infrared excess and on their particular mid-infrared shells emission, using photometric data from the 2MASS and Spitzer/GLIMPSE surveys. Each candidates OGLE light curves were then visually inspected and we selected for spectroscopy follow-up those that underwent large and rapid declines.} 
{We discover two new R Coronae Borealis stars, but also indicate four new possible candidates. We emphasize that all of the 7 known RCB stars located in both the Spitzer/GLIMPSE and OGLE-III fields were also re-discovered, which illustrates the high efficiency of our analysis.} 
{The proposed new technique to find RCB stars has been successful. It can now be extended to larger areas, in particular where the interstellar extinction is too high to be monitored by optical microlensing surveys, such as the inner part of the Galactic bulge.}


\authorrunning{P. Tisserand et al.}
\titlerunning{New R Coronae Borealis stars discovered in OGLE-III Galactic bulge fields}

\maketitle

\section{Introduction \label{sec_intro}}

R Coronae Borealis star (RCB) is a rare type of evolved H-deficient and carbon-rich supergiant star that is increasingly suspected to result from the merger of two white dwarfs (one CO- and one He-), called the Double Degenerate (DD) scenario. The DD model has been strongly supported by two observations: an $^{18}$O over abundance in seven H-deficient carbon and cool RCB stars \citep{2007ApJ...662.1220C}, and the large abundances of fluorine found in the hotter RCB atmospheres \citep{2008ApJ...674.1068P}. In this scenario, observations of RCB stars would not only help us to constrain simulations of low mass Double Degenerate merging events \citep{2008ASPC..391..335F,2008arXiv0811.4646D}, but RCB birthrates would also help us to constrain the rate of white dwarf mergers that could become, at higher mass, supernovae type Ia progenitors \citep{1984ApJ...277..355W,2005ApJ...629..915B}.

R CrB is the prototype RCB star. Over the past $\sim$200 years, it has displayed unpredictably fast declines in brightness, up to 9 magnitudes in the visible. These declines are the main signature of RCB type stars. They are caused by newly formed dust clouds that shadow their central stars. These clouds, consisting of amorphous carbon, are understood to be formed near the stellar atmosphere and accelerated away by radiation pressure. A detailed review of RCB characteristics was given by \citet{1996PASP..108..225C}.

RCB stars in the Large Magellanic Cloud are distributed mostly along the bar \citep{2001ApJ...554..298A,2009A&A...501..985T,2009AcA....59..335S}. Reciprocally, the majority of Galactic RCB stars seem to be concentrated in the bulge. Indeed, a high number of RCB stars (13) have been found inside the Galactic bulge \citep{2008A&A...481..673T} with the surprising peculiarity of being distributed in a thin disk structure. These new discoveries indicate that the density of these rare stars is expected to be high at low Galactic latitude (more than 1 per deg$^2$ at $\mid$b$\mid\leqslant$1 deg).

It is necessary to increase the number of known bulge RCB stars to understand their spatial distribution and therefore constrain their age and past evolution. However this research is difficult in this particular area of the sky because of the high interstellar extinction and the very high density of stars. We note that with an RCB phase lifetime of about $10^5$ years, as predicted by theoretical evolution models \citep{2002MNRAS.333..121S}, and an estimated He-CO white dwarfs merger birthrate between $\sim10^{-3}$ and $\sim5\times10^{-3}$ per year \citep{2001A&A...365..491N,2009ApJ...699.2026R}, we can expect between 100 and 500 RCB stars to exist in our Galaxy. We actually know about 50 of them (see \citet{2008A&A...481..673T}, and references therein), which is only a factor of two more than the Magellanic Clouds, where 23 RCBs are known \citep{2001ApJ...554..298A,2004A&A...424..245T,2009A&A...501..985T}.

In this article, we present a new method to search for RCB stars at low Galactic latitude in the OGLE-III fields. This analysis uses pragmatic criteria and should be considered as a test before extending an RCB search to the entire bulge area. We used the RCB stars peculiar circumstellar shell emission \citep{2009A&A...501..985T} to reduce the number of preselected candidates for further study, instead of highly time-consuming classical methods based on optical light curve analysis. Thanks to the publically available mid-infrared Spitzer/GLIMPSE and near-infrared 2MASS databases, only a few hundred objects were preselected before visual inspection of their OGLE-II and -III light curves.

This article reports on the analysis and discovery of two new RCB stars and four new candidates. The photometric and spectroscopic data used are presented in Sect.~\ref{sec_obs} and the detection techniques in Sect.~\ref{sec_mining}. The general characteristics of the newly discovered RCBs are discussed in Sect.~\ref{sec_newRCB}.


\section{Observational data \label{sec_obs}}

Our analysis used photometric data from four different surveys: 2MASS, Spitzer/GLIMPSE, OGLE-II/III, and EROS-2. The last two projects monitored millions of stars distributed over about one hundred square degrees in the Magellanic Clouds and the Galactic bulge to search primarily for microlensing events produced by potential dark matter candidates in the halo \citep{2007A&A...469..387T,2009MNRAS.397.1228W} or to ordinary stars in the Galactic plane \citep{2006ApJ...636..240S,2006A&A...454..185H}. The OGLE (Optical Gravitational Lensing Experiment) survey has known multiple phases. In this article, we used data from the second, OGLE-II \citep{2005AcA....55...43S}, and third generation OGLE-III  \citep{2003AcA....53..291U}. OGLE-III started in June 2001 and continued observing until May 2009, when it was upgraded to OGLE-IV. It used the 1.3-m Warsaw telescope located at Las Campanas Observatory, Chile, and a 35$\times$35.5 arc-minute field of view camera that used eight SITe 2048$\times$4096 CCD detectors with 15 $\mu$m pixels resulting in a 0.26 arc-seconds/pixel scale. The OGLE observations were taken mostly in I band, but with occasional V band exposures. The EROS-2 (Experience de Recherche d'Objets Sombres) experiment performed observations between July 1996 and February 2003 using the 1-metre MARLY telescope at ESO La Silla Observatory, Chile. Two wide-field cameras of $\sim$0.95 deg$^2$ field of view were operated and placed behind a dichroic cube, which split the light beam into two broad passbands ($B_e$ and $R_e$). Each camera had a mosaic of eight 2048 $\times$ 2048 CCDs with a pixel size of 0.6$\arcsec$ on the sky.

The near- and mid- infrared photometric data were obtained, respectively, from the 2MASS \citep{2006AJ....131.1163S} and Spitzer/GLIMPSE \citep{2009PASP..121..213C} surveys. 2MASS used two highly-automated 1.3-m telescopes, one at Mt. Hopkins, AZ, and one at CTIO, Chile. Each telescope was equipped with a three-channel camera, each channel consisting of a 256$\times$256 array of HgCdTe detectors, capable of observing the sky simultaneously at J (1.25 microns), H (1.65 microns), and K (2.17 microns). GLIMPSE consists of three separate surveys that observed the Galactic disk and bulge ($\vert$b$\vert\leq$4.5 deg) at wavelengths 3.6, 4.5, 5.8, and 8.0 $\mu$m using the Infrared Array Camera IRAC \citep{2004ApJS..154...10F} on board of the Spitzer satellite.

Spectroscopy of RCB candidates was performed with the Wide Field Spectrograph (WiFeS) instrument \citep{2007Ap&SS.310..255D} attached to the ANU\footnote{Australian National University} 2.3 m diameter telescope at Siding Spring Observatory. WiFes is an integral field spectrograph permanently mounted at the Nasmyth A focus. It provides a $25\times38$ arcsec field with 0.5 arcsec sampling along each of twenty five $38\times1$ arcsec slitlets. The visible wavelength interval is divided by a dichroic at around 600 nm  feeds two essentially similar spectrographs. Observations are presented, with a 2-pixel resolution of 2 \AA{}.

\section{Search of Galactic bulge RCB stars \label{sec_mining}}
 
We used a new method to identify RCB stars. This method is based on the particular RCB's circumstellar shell brigthness. \citet[Fig.4]{2009A&A...501..985T} illustrated that on average Magellanic RCB shells are fainter at 8 $\mu$m than those of most common AGB stars, but also cooler than those of classical carbon stars. This can be explained by the highly non-spherical dust distribution around RCB stars as their dust shell is formed from the ejection of dust clouds in random directions \citep{1996PASP..108..225C,2004A&A...428L..13D,2009ApJ...707..573W}. A selection of RCB candidates based on their mid-infrared magnitudes and colours should therefore greatly simplify the search for RCB stars as the number of objects selected for further follow-up drops dramatically. Effectively, classical methods are based mainly on optical light curve analyses dedicated to identify random fast decreases in brightness. These methods are highly time-consuming as RCB have to be searched for among millions of light curves. Furthermore, the mid-infrared method is less biased: inactive RCB stars can now be found, as well as very active ones that remain fainter in the optical.

Our search of bulge RCB stars was performed in four consecutive steps. The details and complications of the selection criteria are described below. First, we selected candidates based on broadband mid-infrared magnitudes and colours using the Spitzer GLIMPSE catalogues (II and III) \citep{2009PASP..121..213C}. Second, we rejected all objects not showing a near-infrared excess in the 2MASS catalogue. Third, we examined at the optical light curves of all remaining objects with a counterpart in the OGLE-II and -III catalogues and selected those objects that exhibit suspicious rapid declines in brightness. Finally, we followed up spectroscopically most of the selected objects.

We note also that we observed spectroscopically a number of optically bright objects selected on the basis only of the two infrared criteria but located outside the OGLE-III fields. A few of them have an atmosphere rich in carbon and are therefore now considered as RCB star candidates. Additional explanations are given in Sect.~\ref{sec_spectro}.

\subsection{Selection from near-infrared and mid-infrared properties}

The Spitzer GLIMPSE surveys observed the Galactic bulge and plane, and catalogued millions of objects \citep{2009PASP..121..213C}. The published archive catalogues contain four mid-infrared broadband magnitudes centred on 3.6, 4.5, 5.8, and 8.0 $\mu$m, obtained with the Spitzer IRAC camera. Despite the short exposure time used (1.2 s), we estimated that bulge RCB stars should be saturated in the first two bands and be at the limit of saturation in the last two\footnote{Saturation limit : [3.6]$_{sat}\sim$~7.0 ; [4.5]$_{sat}\sim$~6.5 ;  [5.8]$_{sat}\sim$~4.0 ; [8.0]$_{sat}\sim$~4.0 \citep{2007AJ....134.2099R}}. We therefore only used the broadband magnitudes [5.8] and [8.0] to select RCB star candidates. In Table~\ref{tab.RCB.MidIR}, we list the mid-infrared magnitudes of the known bulge RCB stars found in the GLIMPSE archive catalogues. We used these archive catalogues because they are more complete and therefore optimise our chance of finding RCB stars that are even saturated. We note that the saturation level depends on the background emission, which experiences strong variation in the GLIMPSE fields. We obtained measurements for all the 13 known bulge RCB stars present in the GLIMPSE fields, but one, EROS2-CG-RCB-1, has no measurement in the [8.0] band. It is also the closest known RCB to the Galactic plane, where the background emission is higher (see \citet{2009PASP..121..213C} for more explanation).

We used three different selection criteria to optimise our chance of finding bulge RCB stars. Each criterion uses the near-infrared 2MASS J, H, and K magnitudes, as well as the Spitzer/GLIMPSE [5.8] and [8.0] ones, which are explained hereafter. The first criterion is described in more detail as it corresponds to the favourable case where all magnitude measurements are available. The numbers of selected objects from each selection criterion are listed in Table~\ref{tab.OBJ.preselect}.

\begin{table} 
\caption{Numbers of objects preselected.
\label{tab.OBJ.preselect}}
\medskip
\centering
\begin{tabular}{lccccc}
\hline
Criteria :  & 1 & 2 & 3 & total\\
\hline
Entire GLIMPSE-II & 772 & 136 & 1072 & 1980\\
Entire GLIMPSE-III & 625 & 204 & 299 & 1128\\
Located in an OGLE-III fields & 154 & 26 & 91 & 271\\
Catalogued by OGLE-III & 150 & 26 & 90 & 266\\
Number of Known RCB found & 5 & 1 & 1 & 7\\
Number of new RCB found & 3 & 0 & 0 & 3\\
\hline
\end{tabular}
\end{table}

\begin{table} 
\caption{Known Galactic RCB stars located in the OGLE-III fields.
\label{tab.known}}
\medskip
\centering
\begin{tabular}{lcc}
\hline
RCB name  & OGLE-III id. & located in a \\
 & & GLIMPSE field ?\\
\hline
EROS2-CG-RCB-1 &   BLG195.5.23   &  yes \\
EROS2-CG-RCB-3 &   BLG182.1.96290  & yes \\
EROS2-CG-RCB-5 &   BLG138.1.130117  &  yes \\
EROS2-CG-RCB-10 &  BLG345.6.159428  & yes \\
EROS2-CG-RCB-11 &  BLG343.6.122070  & yes \\
EROS2-CG-RCB-13 &  BLG224.3.130650  & yes \\
EROS2-CG-RCB-14 &  BLG228.2.23975 & no \\
MACHO-135.27132.51 & BLG213.4.129296 & no \\
MACHO-401.48170.2237 & BLG214.2.215079 & yes \\
V 517 Oph & x & no \\
V 1783 Sgr  & BLG160.8.4580 & no \\
V 739 Sgr & BLG193.1.103298 & no \\
V 3795 Sgr & BLG252.6.166190 & no \\
VZ Sgr & x & no \\
GU Sgr & BLG275.2.39337 & no \\
\hline
\multicolumn{3}{l}{x : Not catalogued as too bright on template image.}\\
\hline
\end{tabular}
\end{table}

\begin{figure*} 
\includegraphics[scale=0.7]{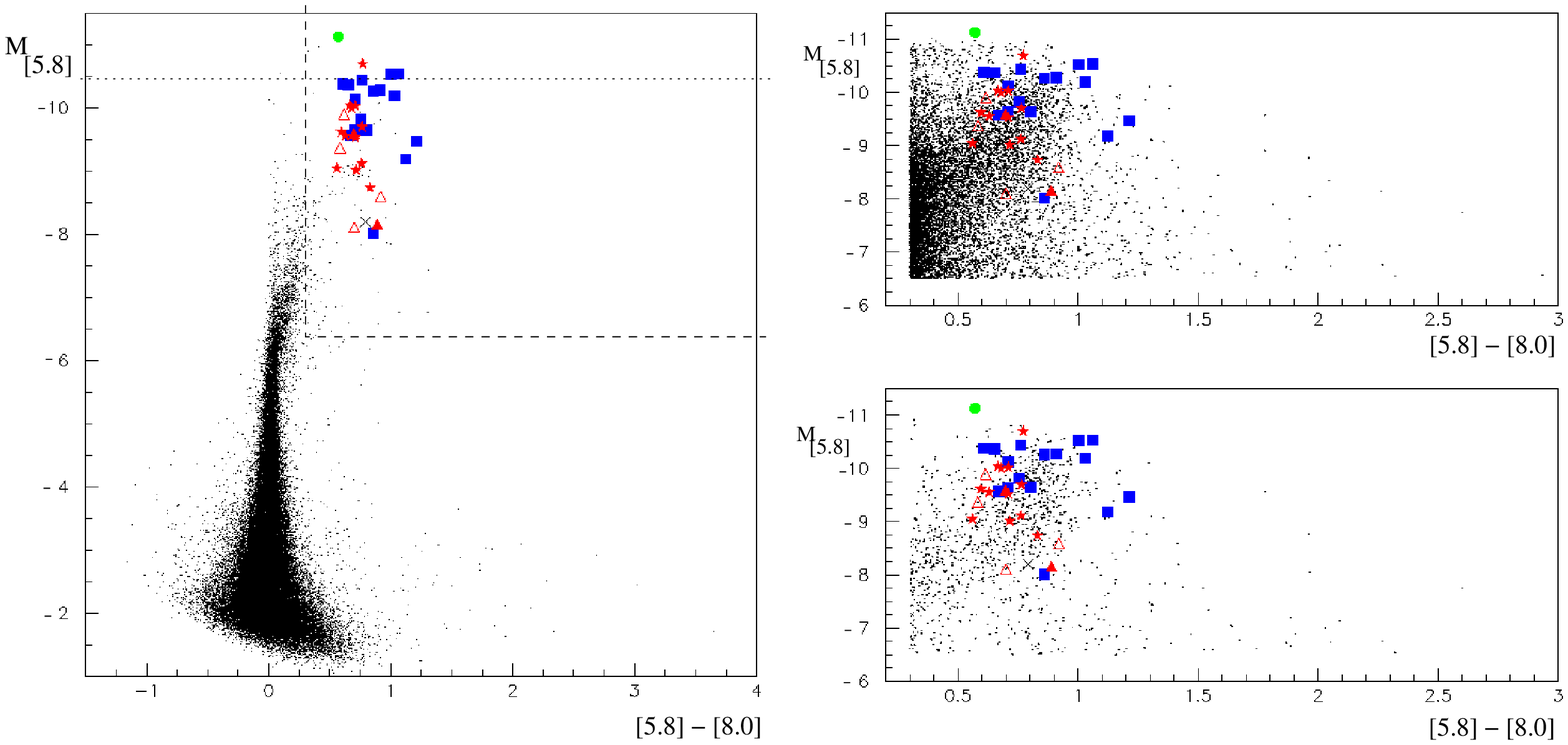}
\caption{\textit{Left} : CMD absolute magnitude $M_{[5.8]}$ versus $[5.8] -[8.0]$. For the sake of clarity, the dot points represent only all objects catalogued in one 1x2.5 deg$^2$ GLIMPSE area centred on $l\sim0.5$, $b\sim-3$ deg. The known RCB stars are represented by larger symbols : blue squares for the Large Magellanic Cloud \citep{2001ApJ...554..298A,2009A&A...501..985T}, green circle for the Small Magellanic Cloud \citep{2004A&A...424..245T}, and red stars for the Galactic bulge \citep{2005AJ....130.2293Z,2008A&A...481..673T}. The newly confirmed bulge RCBs are represented by full red triangles and the new candidates by empty red triangles. The RCB candidate KDM 5651 \citep{2003MNRAS.344..325M} is represented by a cross. We used distance moduli of 18.5, 18.9, and 14.4 for the LMC, SMC, and Galactic center (GC) respectively. The selection area is delimited by dashed lines and the saturation level by a dotted one.
\textit{Right} : Same as left, but the dot points represent all objects selected in the entire GLIMPSE area after the first mid-infrared selection (top) and the remaining objects after the second selection on near-infrared excess (bottom). }
\label{m5-8_fig}
\end{figure*}

\begin{figure*} 
\includegraphics[scale=0.5]{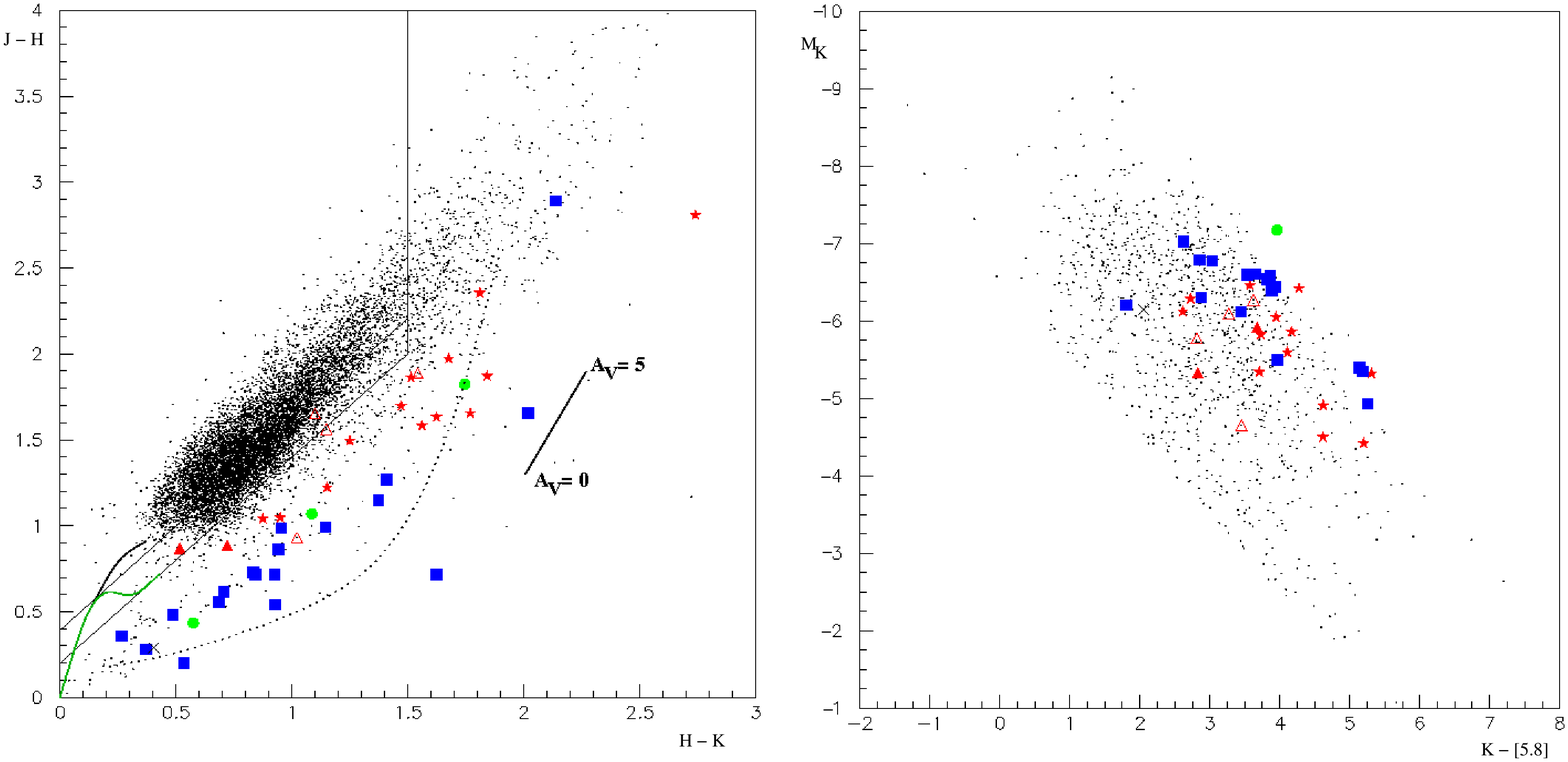}
\caption{\textit{Left}: $J-H$ versus $H-K$ colour diagram. The dot points represent all objects selected with the mid-infrared selection criteria. All the new and known Magellanic and Galactic bulge RCB stars are represented; their symbols are identical to those in Fig.~\ref{m5-8_fig}. The dotted curve corresponds to the combination of black bodies consisting of a 5500 K star and a 1000 K dust shell in various proportions ranging from all 'star' to all 'shell' \citep[from][]{1997MNRAS.285..339F}. The straight lines delimit the near-infrared selection area (see Eqs~\ref{eq.jhk.select1} and \ref{eq.jhk.select2}). The line on the right side represents the reddening vector of \citet{1985ApJ...288..618R}. Also shown are the expected positions (lines in the bottom-left side) of common dwarf (green) and giant (black) stars from \citet{1988PASP..100.1134B}. \textit{Right} : CMD absolute magnitude $M_K$ versus $K-[5.8]$ of all objects that passed the two infrared selection criteria.}
\label{jhk_fig}
\end{figure*}

\begin{itemize}
\item \textit{Criterion 1} : Figure~\ref{m5-8_fig}, left side, presents the colour magnitude diagram (CMD), i.e., plot of absolute magnitude $M_{[5.8]}$ versus $[5.8] -[8.0]$, of all objects located in one GLIMPSE-III field and presenting non-null values in [5.8] and [8.0] magnitudes. The known Galactic bulge and Magellanic RCBs are indicated as well as the selection limits and the saturation level. We selected all objects brighter than $M_{[5.8]}<-6.5$ with a colour index $[5.8]-[8.0]$ redder than 0.3. The selection criteria were based on the known Magellanic and Galactic bulge RCB magnitudes and colours. We can conclude on the basis of resuls for the Magellanic RCBs that most bulge RCBs should not be saturated in the [5.8] band. However, as RCBs have a positive $[5.8]-[8.0]$ colour and the saturation level in these two bands is identical, we expect that some bulge RCBs be saturated in the [8.0] band\footnote{Note that it should not affect our selection criteria if a [8.0] measurement exists.}. This last remark is supported by the almost constant $[5.8] -[8.0]$ colour observed for the bulge RCBs. Finally, we note that there is good overlap between the Magellanic and Galactic bulge RCBs, supporting again a bulge location for these last ones.

The two figures on the right side of Fig.~\ref{m5-8_fig} present all selected objects in the GLIMPSE-III fields after the mid-infrared selection (top figure) that we have just described and the remaining objects after the near-infrared second selection (bottom one). This near-infrared selection was designed to select all objects displaying a near-infrared excess. The applied near-infrared criteria are shown in Fig.~\ref{jhk_fig}, on the left side. We retained only all objects with measurements in all three 2MASS bands that pass the following pragmatic criteria depending on the GLIMPSE survey analysed, GLIMPSE-II (GII) area being more affected by interstellar extinction than GLIMPSE-III (GIII):

\begin{equation}
$$ GII  : J-H < 1.2 (H-K) + 0.4\hspace{2 mm}or\hspace{2 mm} H-K > 1.5 $$
\label{eq.jhk.select1}
\end{equation}

\begin{equation}
$$ GIII : J-H < 1.2 (H-K) + 0.2\hspace{2 mm}or\hspace{2 mm} H-K > 1.5 $$
\label{eq.jhk.select2}
\end{equation}

We note that this near-infrared selection criteria reject most of the classical carbon stars, distributed along the main locus of Fig.~\ref{jhk_fig}, left side. The right-hand side shows the distribution of all remaining stars after both infrared selection criteria in a CMD of absolute magnitude $M_K$ vs. $K-[5.8]$. We interpret the $\sim$0.8 magnitude difference between Magellanic and Galactic bulge RCBs as being caused by the interstellar extinction that affects the latter group.

Interestingly, we emphasize that an overdensity appeared after both infrared selections had been applied, in the colour magnitude diagram $[5.8]$ vs. $[5.8]-[8.0]$ (Fig.~\ref{m5-8_fig}, bottom - right) at the expected brightness and temperature of known RCB star shells.

\item \textit{Criterion 2} : The number of RCB stars with no information in the mid-infrared [8.0] band due to saturation, such as EROS-2-CG-RCB-1, may be important. This situation may be more common closer to the Galactic plane as the background emission becomes higher. Therefore, for this second series of criteria, we kept all objects that have no measurement in the [8.0] band but a valid one in [5.8], and selected all bright objects with $M_{[5.8]}<-8.5$. We increased the value of this last selection level in comparison to the first criterion because only really bright RCBs in [5.8] should be saturated in the [8.0] band (RCBs $[5.8]-[8.0]$ colour index values range between 0.5 and 1.5 (see Fig.~\ref{m5-8_fig})). Afterward, we applied the same near-infrared selection cuts as presented in criterion 1 (see Eqs~\ref{eq.jhk.select1} and \ref{eq.jhk.select2}), except that we added an upper limit to the K band brightness of $M_K<-8$.

\item \textit{Criterion 3} : RCB stars could have been observed by 2MASS during an obscured phase caused by dust clouds. Therefore, we can expect to find RCB stars with missing information in near-infrared but consistent values in the mid-infrared. This problem should also be more common at lower latitude as the interstellar extinction becomes higher. We kept all objects with a valid measurement in K, [5.8], and [8.0] but missing one in J and/or H. We applied the same mid-infrared selection as for criterion 1 and added two further limits related to the K band of $M_K<-8$ and $K-[5.8]>2.0$. We note that with an $M_K$ absolute magnitude of about -6, RCB stars are almost always visible in that particular band towards the bulge: even in the case of an extreme extinction, $A_V\sim30$, bulge RCB stars would still have a K magnitude of $\sim$11.4 ($A_K\sim0.1 A_V$).


\end{itemize}

Seven previously known RCB stars were in our fields and should have been discovered using the selection criteria that we have just explained. They were all catalogued by both Spitzer/GLIMPSE and OGLE-III surveys and we list them in Table~\ref{tab.known}. We successfully found all of these 7 RCB stars. We note that they would all have been considered directly as RCB stars because they exhibit sharp declines in their respective OGLE light curve\footnote{The OGLE-III light curves of all known RCB stars are available at the following URL : http://ogle.astrouw.edu.pl/ogle3/rcom/rcom.html . This web site present a useful real-time monitoring of known RCBs \citep{2008AcA....58..187U}}.

\begin{table*}
\caption{General information about the newly confirmed and candidate RCB stars.
\label{tab.RCB.GenInfo}}
\medskip
\centering
\begin{tabular}{lllll}
\hline
\hline
RCB name & OGLE-III  & OGLE-II & Coordinates ($J_{2000}$) & Other identifier  \\
     & identifier & identifier &  & \\
\hline
\multicolumn{5}{c}{New RCB stars} \\
\hline
\object{OGLE-GC-RCB-1}  & BLG333.5.145097 & BUL\_SC43.326584  &  17:35:18.125 -26:53:49.17 & NSV 22773 ; Terz 2637\\
 & & & & IRAS 17321-2652 \\
\object{OGLE-GC-RCB-2}  & BLG101.4.230792 & BUL\_SC3.793780  & 17:53:57.078 -29:31:42.90  & IRAS 17507-2931 \\
			 & BLG195.1.8088   &  &   & ASAS\_175357-2931.8  \\
\hline
\multicolumn{5}{c}{New RCB star candidates} \\
\hline
 \object{OGLE-GC-RCB-Cand-1} & BLG194.1.90513   & BUL\_SC37.133492 & 17:52:14.926 -29:38:30.47  &  EROS2-cg0034l11292 \\
 \object{OGLE-GC-RCB-Cand-2}  & BLG148.5.148667 &  	       &  17:51:19.507 -32:43:21.44 & EROS2-cg0766k15255 \\
			& BLG155.8.10916  &  &   & \\
 \object{GLIMPSE-RCB-Cand-1}  &   & & 17:46:45.916 -25:03:14.05  &  \\
 \object{GLIMPSE-RCB-Cand-2}  &   & & 17:57:49.999 -18:25:22.89  &  \\
\hline
\multicolumn{5}{l}{}\\
\end{tabular}
\end{table*}

\begin{table*} 
\caption{Near-IR photometry.
\label{tab.RCB.IRInfo}}
\medskip
\centering
\begin{tabular}{llrrrlrrr}
\hline
\hline
Name & JD Epoch 2MASS & $J_{\mathrm{2MASS}}$ & $H_{\mathrm{2MASS}}$ & $K_{\mathrm{2MASS}}$ & JD Epoch DENIS & $I_{\mathrm{DENIS}}$ & $J_{\mathrm{DENIS}}$ & $K_{\mathrm{DENIS}}$  \\
\hline
& & & New RCB stars  &    &    &  &  &  \\
\hline
OGLE-GC-RCB-1 & 2450996.6585$^{\vee}$ & 10.091 & 9.204 & 8.484 &  2451118.578086$^{\wedge}$ & 11.627 & 9.723 & 8.028 \\
OGLE-GC-RCB-2 & 2451010.6682$^{\Diamond}$ & 10.453 & 9.585 & 9.069 & nd & nd & nd & nd  \\
\hline
\multicolumn{9}{c}{New RCB star candidates}\\
\hline
OGLE-GC-RCB-Cand-1 & 2451010.6575$^{\Diamond}$ & 11.372 & 9.719 & 8.621 & 2451404.576447$^{\Diamond}$ & 15.238 & 11.320 & 8.453 \\
OGLE-GC-RCB-Cand-2 & 2451040.4809$^{\vee}$ & 13.174 & 11.284 & 9.743 &  2451299.886319$^{\vee}$ & nd & 13.404  & 9.702 \\
GLIMPSE-RCB-Cand-1  & 2451825.4976$^{?}$ & 10.249 & 9.319 & 8.299 &  2451101.670528$^{?}$ & 12.539 & 10.173 & 8.236 \\
GLIMPSE-RCB-Cand-2 & 2450978.7733$^{?}$ & 10.837 & 9.274 & 8.127 &  2451764.650718$^{?}$ & 13.656 &  9.956 & 7.689 \\
\hline
\multicolumn{9}{l}{$\star$: during a faint phase, $\Diamond$: during a bright phase, $\vee$ and $\wedge$: during a dimming or recovering phase, ?: phase unknown. }\\
\end{tabular}
\end{table*}

\begin{table*}
\caption{Spitzer/GLIMPSE II and III and MSX magnitudes for bulge RCB stars.
\label{tab.RCB.MidIR}}
\medskip
\centering
\begin{tabular}{lcccccccc}
\hline
\hline
  & & & Spitzer IRAC bands & & & & MSX bands &\\
Name & [3.6] & [4.5] & [5.8] & [8.0] & A (8.26 $\mu$m) & C (12.12 $\mu$m) & D (14.65 $\mu$m) & E (21.41 $\mu$m) \\
& & & & & \\
\hline
\multicolumn{9}{c}{Known RCB stars} \\
\hline
EROS2-CG-RCB-1$^{II}$  & 5.388$^s$  & 4.645$^s$ & 3.765 & nd & 4.13 & 3.68 & 3.44 & nd \\
EROS2-CG-RCB-3  & 6.648$^s$  & 5.164$^s$ & 4.372 & 3.662 &  3.40 & 2.81 & 2.59 & 2.33 \\
EROS2-CG-RCB-4  & 6.712$^s$  & 5.546$^s$ & 4.361 & 3.695 &  4.25 & 3.66 & 3.30 & nd\\
EROS2-CG-RCB-5  & 6.525$^s$  & 5.485$^s$ & 4.780 & 4.185 &  3.66 & 3.02 & 3.02 & nd\\
EROS2-CG-RCB-6  & 7.406$^s$  & nd & 5.664 & 4.832 &  4.34 &   3.67 &   3.30 &  nd \\
EROS2-CG-RCB-7  & 7.119$^s$  & nd & 5.353 & 4.793 &  4.95 & nd & nd & nd \\
EROS2-CG-RCB-8$^{II}$  & 7.019$^s$  & 6.310$^s$ & 5.281 & 4.519 &  4.46 & 3.96 & 3.71 & 3.69\\
EROS2-CG-RCB-9  & 6.605$^s$  & 6.109$^s$ & 4.699 & 3.935 & 3.90  &  3.15  &  2.76  & nd  \\
EROS2-CG-RCB-10 & 5.100$^s$  & 4.594$^s$ & 3.704 & 2.933 &  2.91   &  2.24  &   2.04  &    1.62 \\
EROS2-CG-RCB-11 & 7.021$^s$  & 6.030$^s$ & 5.385 & 4.670 &  4.09   &   3.47   &   3.16   &  nd \\
EROS2-CG-RCB-13 & 6.849$^s$  & 6.122$^s$ & 4.869 & 4.159 &  4.31    &  3.57   &   3.87   &  nd \\
EROS2-CG-RCB-14$^{o}$   & nd & nd & nd & nd & 3.75 &   3.06 &   2.98 &  nd\\
MACHO-301.45783.9$^{o}$ & nd & nd & nd & nd & 5.06 & nd & nd & nd \\
MACHO-308.38099.66  & 6.200$^s$  & 5.541$^s$ & 4.847 & 4.217 & nd & nd & nd & nd  \\
MACHO-401.48170.2237$^{II}$ & 6.702$^s$  &   nd  & 4.404 & 3.724 &  nd & nd & nd & nd  \\
\hline
\multicolumn{9}{c}{Known DYPer type stars} \\
\hline
EROS2-CG-RCB-2$^{II}$  & 8.167  &  7.893  & 7.667  & 7.367  &  nd & nd  & nd  & nd \\
\hline
\multicolumn{9}{c}{New RCB stars} \\
\hline
OGLE-GC-RCB-1  &  6.449$^s$   & 5.932$^s$   & 4.813   & 4.116 & 4.55 &   3.99 &   3.73 & nd  \\
OGLE-GC-RCB-2$^{II}$  &  7.678$^s$   & 6.935$^s$   & 6.243   & 5.354 & 3.94 &   3.02 &   3.01 & 2.43 \\
\hline
\multicolumn{9}{c}{New RCB star candidates} \\
\hline
OGLE-GC-RCB-Cand-1$^{II}$  &  7.602 & 6.751 & 5.806 & 4.887 & 4.97 &   3.92 &   3.53 &    2.89 \\
OGLE-GC-RCB-Cand-2  &  7.552$^s$   & 7.019$^s$   & 6.293   & 5.593 & 5.25 & nd & nd & nd \\
GLIMPSE-RCB-Cand-1  &  6.789$^s$   &  5.822$^s$   &  5.030   &  4.447   &  5.05 & nd &    3.46 &  nd  \\  
GLIMPSE-RCB-Cand-2  &  6.062$^s$   &  5.239$^s$   &  4.505   &  3.890   &  3.45 &   2.76 &   2.68 &  nd  \\ 
\hline
\multicolumn{9}{l}{nd = not detected ; $^s$: probably saturated, measurement can't be trusted. } \\
\multicolumn{9}{l}{$^{II}$ Spitzer magnitudes from GLIMPSE-II-Archive catalogues, instead of GLIMPSE-III-Archive for the others } \\
\multicolumn{9}{l}{$^{o}$ Located outside Spitzer GLIMPSE fields.} \\
\end{tabular}
\end{table*}






\subsection{OGLE light curves}

\begin{figure*} 
\includegraphics[scale=0.7]{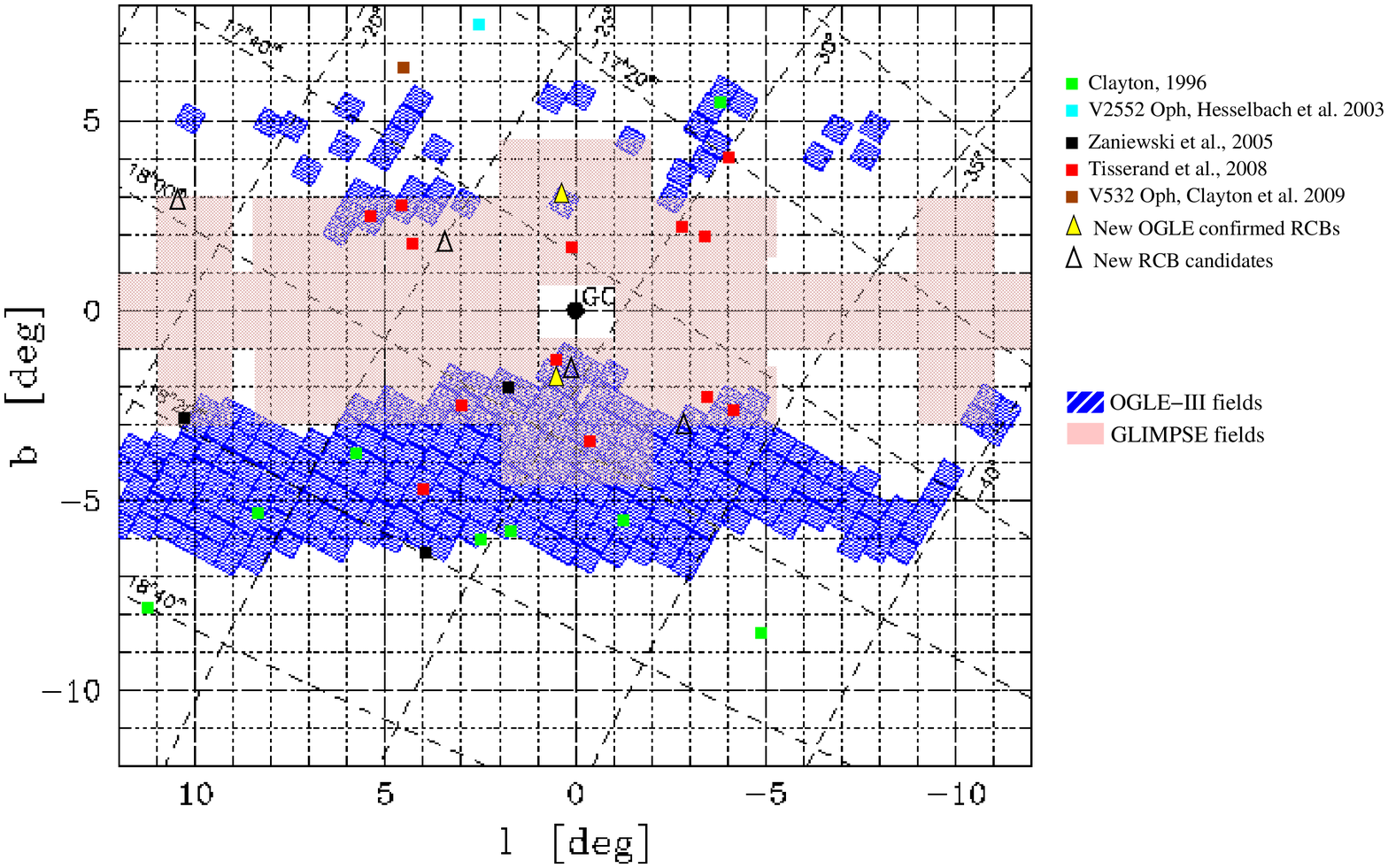}
\caption{Galactic bulge map with a representation of the area observed by the Spitzer/GLIMPSE survey and the fields monitored by the OGLE-III survey. The location of the newly confirmed and candidate Galactic RCBs stars are indicated along with 13 discovered inside the bulge by \citet{2008A&A...481..673T}, four reported by \citet{2005AJ....130.2293Z}, the newly active warm RCB V 2552 Oph \citep{2003PASP..115.1301H}, V 532 Oph \citep{2009PASP..121..461C}, and the other previously known RCBs in this region (V1783 Sgr, V739 Sgr, V3795 Sgr, VZ Sgr, V517 Oph, Gu Sgr, V348 Sgr, and WX Cra). }
\label{fields_fig}
\end{figure*}

Among all selected objects in the GLIMPSE fields, we inspected visually the light curves of the ones catalogued in the OGLE-II and OGLE-III fields. The overlap between the GLIMPSE area and the OGLE-III fields is represented in Fig.~\ref{fields_fig} (Note that the area covered by the OGLE-III fields include the OGLE-II ones). Of the 271 candidates located in the OGLE-III fields, 266 were catalogued. Four of the remaining ones were saturated in the reference image and the other one was too faint\footnote{We note that this particular star (ra = 18:03:20.368  dec = -27:32:24.27) was catalogued in the USNO-B1.0 catalogue in 1982.4 as a bright star : R2$\sim$13.68 . }. The light curve visual inspection left us with no doubt about the RCB nature of two objects, named OGLE-GC-RCB-1 and OGLE-GC-RCB-2. Their coordinates are presented in Table~\ref{tab.RCB.GenInfo}, light curves in Fig.~\ref{lcogle_fig} and ~\ref{lceros_fig}, and charts in Fig.~\ref{chart_fig}. OGLE-GC-RCB-1 exhibits six consecutive drops from 4 to up to 9 magnitudes in its OGLE-II and -III light curves, while OGLE-GC-RCB-2 has multiple rapid and spectacular declines, up to 8 mag., immediately followed by rapid recoveries. 
Overall, among the 266 stars, 20 stars were selected during the visual inspection process : 10 were selected due to non-periodic and sudden declines present in their light curves, and another 10 were selected because their light curves resemble RCB star behaviour during a recovery or a bright phase. We followed-up these 20 stars spectroscopically.

\begin{table} 
\caption{Magnitudes I$_{max}$ and colours (V-I)$_{max}$ of the new OGLE Galactic RCBs at maximum brightness, with derived absolute magnitudes M$_V$ and intrinsec colours (V-I)$_0$ after extinction correction.
\label{tab.mag}}
\medskip
\centering
\begin{tabular}{lcccccc}
\hline
OGLE RCBs & (V-I)$_{max}$ & I$_{max}$ & A$_{Be}$ & V$_0$ & (V-I)$_0$ & M$_V$ \\
\hline
RCB-1 & 3.20 & 11.30 & 2.96 & 10.87 & 1.53 & -3.53 \\
RCB-2 & 2.10 & 11.50 & 3.06 & 9.85 & 0.22 & -4.55 \\
Cand-1 & 5.20 & 14.70 & 3.53 & 15.57 & 3.20 & 1.17 \\
Cand-2 & 5.20 & 15.20 & 3.09 & 16.61 & 3.46 & 2.21 \\
\hline
\end{tabular}
\end{table}


\subsection{Spectra \label{sec_spectro}}

If a well-sampled light curve is available, identification with the RCB class can be made with fairly high confidence because of the distinct nature of the RCB brightness drops (fast and up to 9 mag.). However, if these declines remain small,  spectroscopic information is necessary to reveal and confirm their true nature.

\begin{figure}
\centering
\includegraphics[scale=0.3]{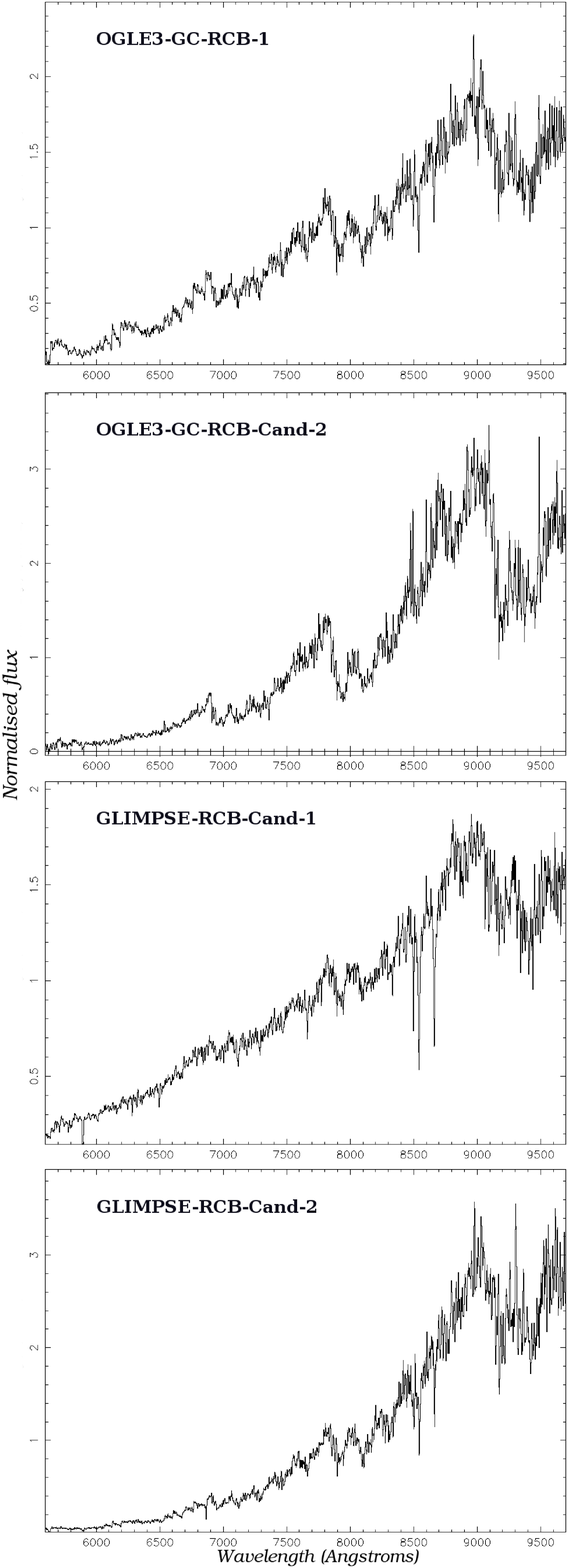}
\caption{Spectra of the new confirmed and candidates RCB stars. The flux (F$_\lambda$) is normalised to the flux at 8000 $\AA{}$.}
\label{spectra_fig}
\end{figure}

Most of the 20 stars selected for spectroscopy follow-up, were found to be either M giant stars exhibiting strong TiO bands features in their spectra, or hot stars presenting emission lines for half of them. However one star, named hereafter OGLE-GC-RCB-1, exhibits strong carbon features in its spectrum and can be assumed to be a new confirmed RCB stars. More discussion about this star is presented in Sect.~\ref{sec_newRCB}. Unfortunately, we did not obtain spectra for OGLE-GC-RCB-2 and another star considered to be a good RCB candidate and named hereafter OGLE-GC-RCB-Cand-1 in Table~\ref{tab.RCB.GenInfo}. The carbon spectra obtained are presented in Fig.~\ref{spectra_fig}.

We also decided to target spectroscopically some optically bright objects selected by one of the 3 infrared criteria, but located outside the OGLE-III fields, to estimate the efficiency of finding RCB stars during a blind survey (i.e without light curves being available). Among the 11 randomly selected GLIMPSE objects, two display an atmosphere rich in carbon (see Fig.~\ref{spectra_fig}) and are therefore now considered as RCB star candidates, they are named GLIMPSE-RCB-Cand-1 and -2 in Table~\ref{tab.RCB.GenInfo}.

From a simple analysis of the spectra, we can empirically compare the different temperatures of the newly confirmed and candidate RCB stars, based on the strength of their Ca II triplet absorption lines. The intensity of these lines is, as shown by \citet{1971ApJ...167..521R}, a good indicator of carbon star temperature: the cooler the temperature, the weaker the lines. The RCB candidate GLIMPSE-RCB-Cand-1 clearly seems to be hotter than the other 3 carbon stars observed and the candidate OGLE-GC-RCB-Cand-2, the coolest. We also searched for the isotope $^{13}C$ in the atmosphere of these stars. We used the spectral atlas of carbon stars compiled by \citet{1996ApJS..105..419B} to identify isotopic $C_2$ and CN bands, respectively at 6100 and 6260 \AA{}. We found no trace of $^{13}C$ in OGLE-GC-RCB-1 and the two candidates GLIMPSE-RCB-Cand-1 and -2, but a positive signature for OGLE-GC-RCB-Cand-2. Indeed, a bandhead feature due to $^{13}$CN is recognisable at $\sim$6260 \AA{}, even if the signal is low in that particular wavelength range.


\section{Discussion about new RCB stars \label{sec_newRCB}}

\begin{figure}
\centering
\includegraphics[scale=0.38]{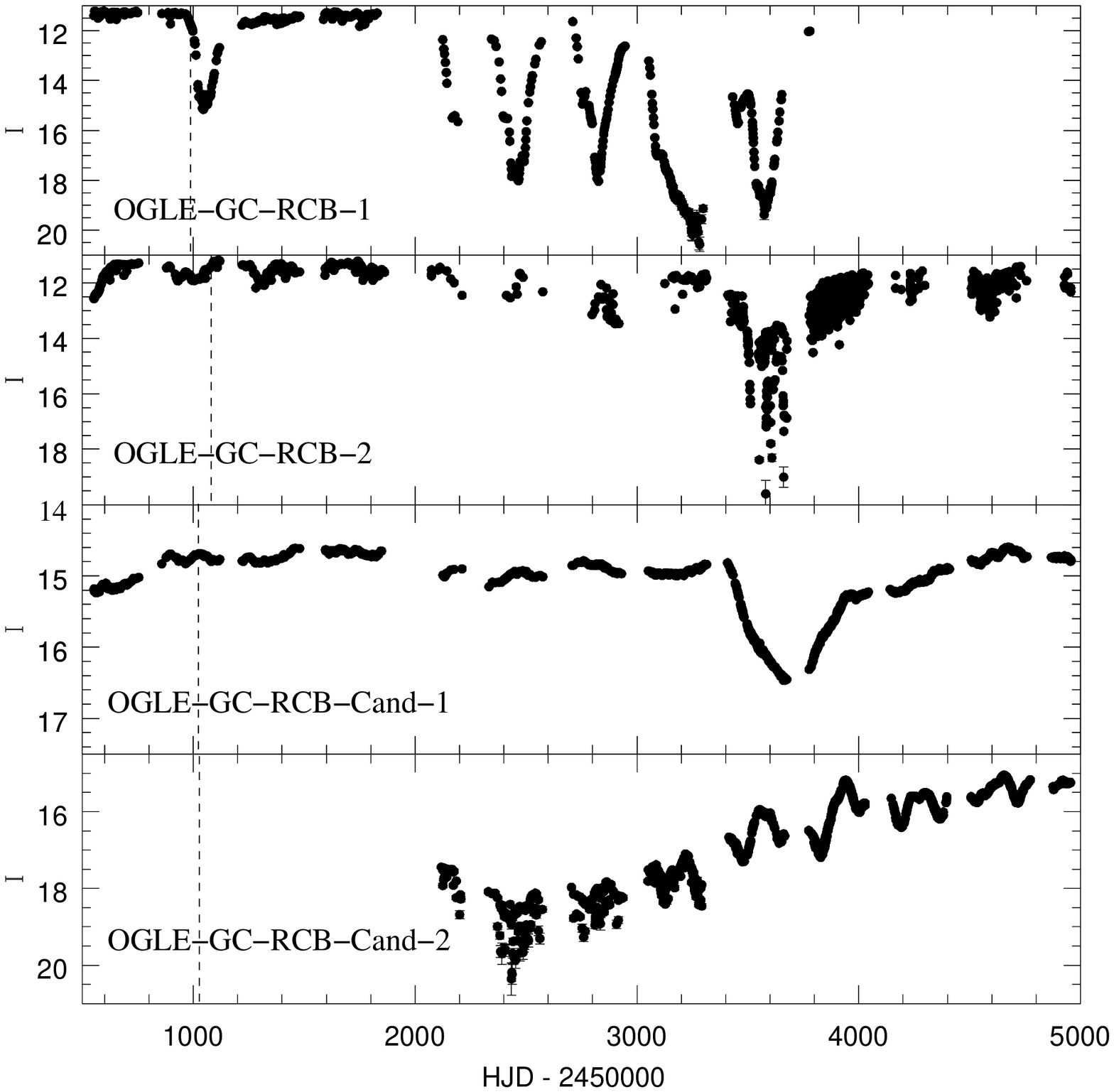}
\caption{OGLE-II and OGLE-III light curves of the new RCB stars, confirmed and candidates. The dashed vertical lines indicate the 2MASS epochs.}
\label{lcogle_fig}
\end{figure}

The newly confirmed and candidate RCB stars are listed in Table~\ref{tab.RCB.GenInfo}, and their near- and mid- infrared photometry in Tables~\ref{tab.RCB.IRInfo} and ~\ref{tab.RCB.MidIR}. Their light curves and charts are presented, respectively, in Figs.~\ref{lcogle_fig} and ~\ref{chart_fig}. 
We note interestingly that we found an entry for most of the known and new RCB stars in the MSX (Midcourse Space Experiment) point source catalogue \citep{2003yCat.5114....0E}. MSX magnitudes are also listed in Table~\ref{tab.RCB.MidIR}. The MSX survey has observed the entire Galactic plane, with $\vert$b$\vert\leqslant$ 6 deg, in four different bands : A (8.26 $\mu$m), C (12.12 $\mu$m), D (14.65 $\mu$m), and E (21.41 $\mu$m). Only objects brighter than A$\lesssim$8 mag were detected, the A band being the most sensitive. We emphasize that MSX magnitudes could help in the future to select new RCB candidates.


In Table~\ref{tab.mag}, we present V and I maximum magnitudes of the 4 OGLE RCBs, confirmed and candidates. Those  magnitudes were estimated from their respective OGLE-II and -III light curves. We applied an interstellar extinction and reddening correction to these magnitudes using the same technique applied in \citet{2008A&A...481..673T}, where an average extinction A$_{Be}$\footnote{Be corresponds to the blue EROS2 filter, see \citet{2007A&A...469..387T}.} was calculated using red clump giant stars located around each RCB. We therefore obtained the extinction-corrected RCBs magnitude V$_0$ and colour (V-I)$_0$ at maximum brightness. Absolute magnitudes M$_V$ were calculated using a bulge distance moduli of 14.4 mag. The values presented in Table~\ref{tab.mag} can be compared to the M$_V$ vs. (V-I)$_0$ colour magnitude diagram presented in \citet[Fig.7]{2008A&A...481..673T}: one can see that OGLE-GC-RCB-2 is a rare warm RCB star, as warm as W Men in the LMC ; OGLE-GC-RCB-1 is a more classical RCB with a temperature around T$_{eff}\sim$5000 K; and OGLE-GC-RCB-Cand-1 and -2 would be extremely cool objects in the RCB class.

We discuss now each new RCB star. We note that none of them have been catalogued in the MACHO database, but two, candidates OGLE-GC-RCB-Cand-1 and -2, were monitored by the EROS-2 survey \footnote{OGLE-GC-RCB-1 and OGLE-GC-RCB-2 were saturated on the EROS2 reference images.}.

\begin{figure}
\centering
\includegraphics[scale=0.38]{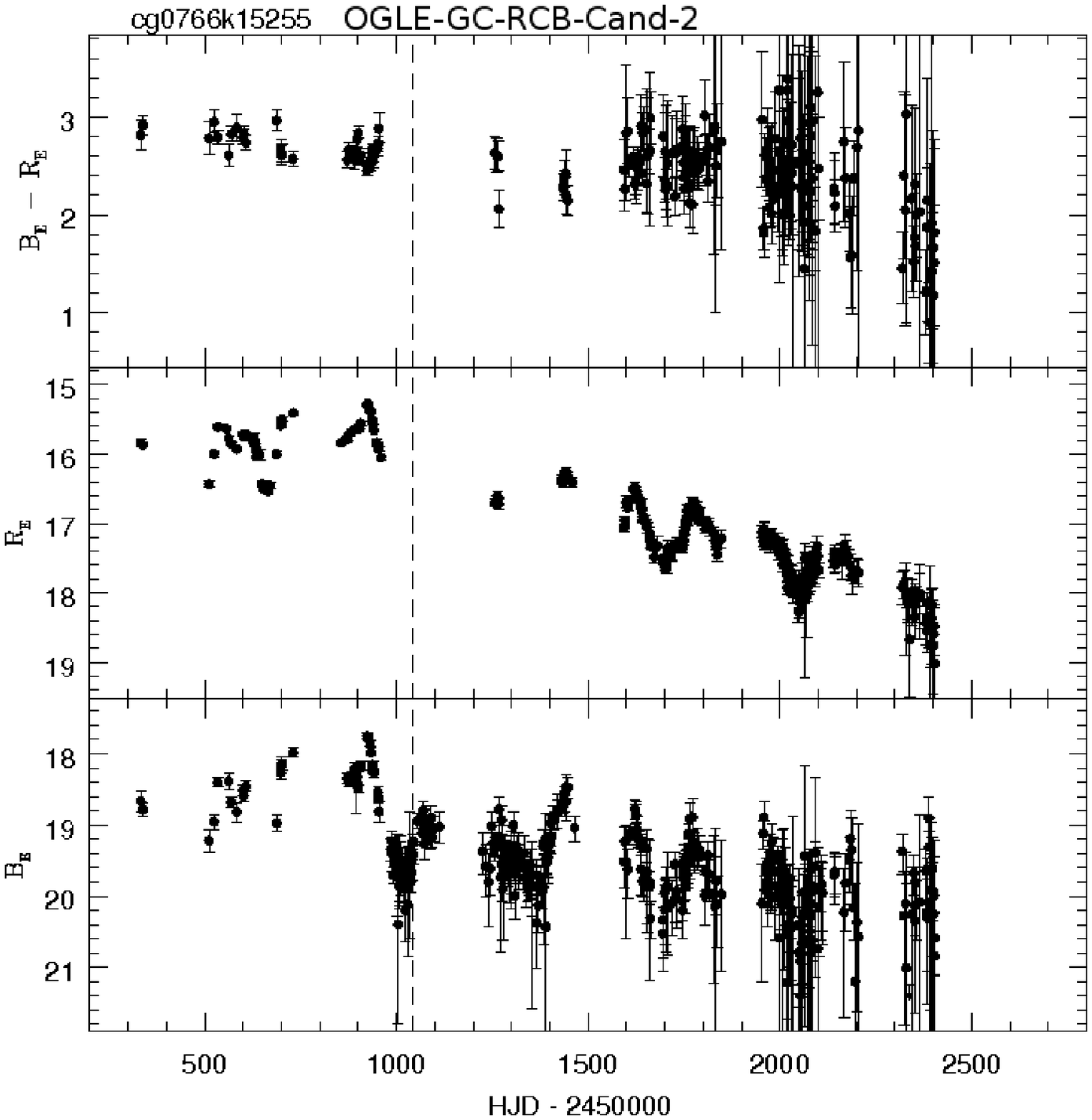}
\caption{EROS-2 light curve of candidate OGLE-GC-RCB-Cand-2 ; top: $B_E-R_E$ colour versus time; middle: $R_E$ light curve; bottom: $B_E$ light curve. The dashed vertical lines indicate the 2MASS epochs.}
\label{lceros_fig}
\end{figure}

\begin{figure}
\centering
\includegraphics[scale=0.55]{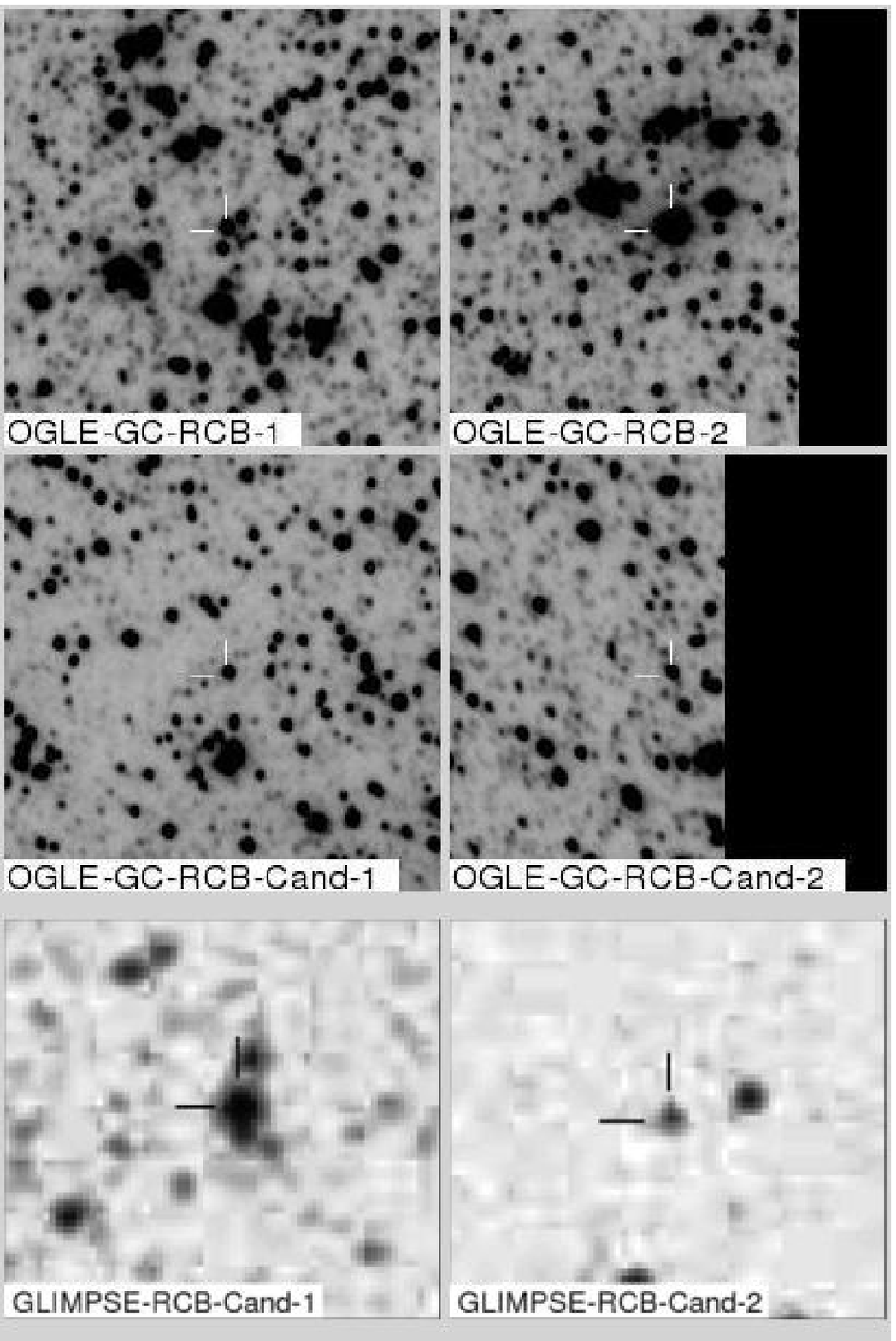}
\caption{Charts of the newly confirmed and candidate RCB stars (1'x1'). North is up, east is to the left. The top images are from the OGLE-III survey, the bottom ones from USNO.}
\label{chart_fig}
\end{figure}

\begin{itemize}

\item OGLE-GC-RCB-1: OGLE-II and -III light curves show that this RCB star has recently experienced a high level of dust production activity as many declines from 4 and up to 9 magnitudes are observable. It was named Terz V 2637 by \citet{1991A&AS...90..451T} due to a four magnitude variation observed in 1987 between two R band epochs separated by 1 month. We also note that it was catalogued as a possible RCB star in the AAVSO International Variable Star Index \citep{2006SASS...25...47W}. With its numerous significant and rapid optical declines, bright shell, carbon spectrum with absence of $^{13}C$ in it, OGLE-GC-RCB-1 is clearly a typical RCB star with a temperature around T$_{eff}\sim$5000 K.

\item OGLE-GC-RCB-2 has undergone a 350 days long phase where multiple rapid and spectacular declines, up to 8 mag., were immediately followed by rapid recoveries. We did not obtain any spectrum of this star, but its light curve and infrared emission indicate that OGLE-GC-RCB-2 is an RCB star. It was bright enough to be monitored by the ASAS-3 survey\footnote{ASAS: All Sky Automated Survey \citep{1997AcA....47..467P}, URL: http://www.astrouw.edu.pl/asas/?page=main}. We underline that OGLE-GC-RCB-2 exhibits the faintest shell of any RCB star. Its mid-infrared brightness is similar to those of both the warm LMC RCB star W Men and the cool LMC RCB candidate KDM-5651 \citep{2003MNRAS.344..325M}. This indicates that OGLE-GC-RCB-2 is in a low ejection activity phase. OGLE-GC-RCB-2 is the warmest RCB star known in the bulge, with an absolute magnitude and intrinsic (V-I)$_0$ colour similar to W Men (see Table~\ref{tab.mag} and \citet[Fig.7]{2008A&A...481..673T}). We note that the temperature of W Men was estimated at $T_{eff}\sim$7000 K by \cite{1990MNRAS.245..119G}.

\item RCB candidate OGLE-GC-RCB-Cand-1 exhibits only a slow $\sim$1.6 mag decline in its OGLE light curve, with a decline rate of $\sim$0.006 mag/day followed by a relatively symmetric immediate recovery, resembling those observed in DY Per type of stars (see \citet{2009A&A...501..985T} for some examples). This last remark is also supported by its position in the $J-H$ versus $H-K$ colour-colour diagram (Fig.~\ref{jhk_fig}), close to the expected locus of classical carbon stars. However we note that OGLE-GC-RCB-Cand-1 was catalogued in the USNO-B1.0 catalogue with a faint magnitude of R$\sim$19.4 mag observed in 1986.4 \citep{2003AJ....125..984M}, which is about 4 magnitudes fainter than the maximum brightness observed by OGLE. Therefore, OGLE-GC-RCB-Cand-1 should still be considered as an RCB candidate. Unfortunatly, we did not obtain any spectrum of that star. From its derived intrinsic (V-I)$_0$ colour, OGLE-GC-RCB-Cand-1 is a cool RCB star candidate.

\item RCB candidate OGLE-GC-RCB-Cand-2 is a special case. Indeed, its OGLE-III light curve diplays a slow recovery phase with large oscillations following a slow decline that lasted more than 2000 days. This slow decline phase is observable in the EROS-2 light curve (Fig. ~\ref{lceros_fig}). On top of the slow decline, one can observe a fast drop of about 2 magnitudes (around HJD$\sim$930) with a decline rate of $\sim$0.026 mag/day. This uncommon optical behaviour resembles that of the RCB star EROS2-LMC-RCB-2 \citep{2009A&A...501..985T} where a slow decline and recovery were also observed\footnote{We note that such declines were also observed in the 200 years light curve of R CrB.}. We note that OGLE-GC-RCB-Cand-2 seems to be a very cool carbon star because its spectrum displays weak Ca II triplet absorption lines. This is consistent with its very red (V-I)$_0$ colour (see Table~\ref{tab.mag}). With this (V-I)$_0$ colour and absolute magnitude M$_V$, OGLE-GC-RCB-Cand-2 may indeed correspond to the faintest and coolest RCB star (see \citet[Fig.7]{2008A&A...481..673T}). However, its derived magnitudes should not be fully trusted as we did not observe with certainty its maximum magnitude. The time spent at maximum brightness during the OGLE-III and EROS-2 observations was short. The low decline rate observed and its relatively faint shell imply that OGLE-GC-RCB-Cand-2 represents a possible link between RCB stars and DY Per type of stars. This last remark is also supported by the non-negligable presence of $^{13}C$ in its atmosphere.

\item RCB candidates GLIMPSE-RCB-Cand-1 and -2 were both found by a spectroscopic follow-up of optically bright objects (R$\sim$14 and $\sim$17, respectively) randomly selected on the basis of their near- and mid- infrared emission. There is no optical light curve available. Both stars have a carbon-rich atmosphere and near- and mid- infrared emissions that are typical of RCB stars. We note that we do not detect any traces of $^{13}C$ features in their spectra, and from the strength of its Ca II triplet absorption lines, GLIMPSE-RCB-Cand-1 is a candidate to be a warm RCB star. For GLIMPSE-RCB-Cand-2, we note a difference of $\sim$0.9 and $\sim$0.4 mag between the two epochs of Denis and 2MASS in the J and K bands, respectively, taken $\sim$786 days apart. This difference seems to indicate that GLIMPSE-RCB-Cand-2 was in an obscured phase during the Denis epoch. Furthermore, we note that GLIMPSE-RCB-Cand-2 is clearly a variable star, as it appears relatively bright on the USNO SERC photographic plate (see its chart, Fig.~\ref{chart_fig}), and is not visible on the USNO POSS I picture. 

\end{itemize}

\section{Summary}

 We have tested new RCB selection criteria to reduce the number of preselected objects for subsequent visual inspection of OGLE light curves prior to spectroscopy follow-up. Thanks to the mid-infrared Spitzer/GLIMPSE and near-infrared 2MASS databases, we have selected 266 stars located in both the Spitzer/GLIMPSE and OGLE-III fields. The analysis has led to the discovery of 2 new RCB stars and 4 new candidates. We have demonstrated that we do not need to use the laborious and time-consuming technique of light curve analysis to find new RCB stars. Furthermore, we rediscovered all of the 7 known RCB stars that were located in both the Spitzer/GLIMPSE and OGLE-III monitored area. This indicates the high efficiency of our analysis.
 
 This new technique can now be used to find RCB stars in areas not monitored optically. RCB star searches can therefore be more easily applied to the entire bulge area, even at low Galactic latitude where interstellar extinction is high. We tested this idea by observing spectroscopically 11 optically bright stars randomly selected on the basis of their mid- and near- infrared emissions and located outside any microlensing survey field. We found that 2 stars exhibit signs of an atmosphere rich in carbon, which may now both be considered as RCB candidates. We emphasize that our chances of finding classical carbon rich stars with our analysis is low as the Galactic bulge is known to have a low density of classical carbon stars, if any \citep{1983AJ.....88.1442B}, and our near and mid - infrared selection eliminates the vast majority of classical carbon stars.





\begin{acknowledgements}
 We thank Tony Martin-Jones and the anonymous referee for their careful reading and comments. We also thank Mike Bessell and Peter Wood for their expertise concerning the spectroscopic data reduction. This work is based in part on observations made with the Spitzer Space Telescope, which is operated by the Jet Propulsion Laboratory, California Institute of Technology under a contract with NASA. The OGLE project is partially supported by the Polish MNiSW grant N20303032/4275. This publication makes use of data products from the Two Micron All Sky Survey, which is a joint project of the University of Massachusetts and the Infrared Processing and Analysis Centre, California Institute of Technology, funded by the National Aeronautics and Space Administration and the National Science Foundation. The DENIS data have also been used. DENIS is the result of a joint effort involving human and financial contributions of several Institutes mostly located in Europe. It has been supported financially mainly by the French Institut National des Sciences de l'Univers, CNRS, and French Education Ministry, the European Southern Observatory, the State of Baden-Wuerttemberg, and the European Commission under networks of the SCIENCE and Human Capital and Mobility programs, the Landessternwarte, Heidelberg and Institut d'Astrophysique de Paris.
\end{acknowledgements}

\bibliographystyle{aa}
\bibliography{RCB-OGLE3-Bulge}


\end{document}